\documentclass[twocolumn,prb,showpacs]{revtex4}
\usepackage{graphicx,psfrag}
\usepackage{slashed,amssymb,amsmath}
\def\no{\noindent}

\newcommand{\nn}{\nonumber}

\begin{document}

\title{Corrections to the self-consistent Born approximation for Weyl fermions}
\author{A. Sinner and K. Ziegler}
\affiliation{
Institut f\"ur Physik, Universit\"at Augsburg,
D-86135 Augsburg, Germany}
\date{\today}

\begin{abstract}

The average density of states of two- and three-dimensional Weyl fermions is studied
in the self-consistent Born approximation (SCBA) and its corrections. The latter have been organized 
in terms of a 1/N expansion. It turns out that an expansion in terms of the disorder strength
is not applicable, as previously mentioned by other authors. Nevertheless, the 1/N expansion
provides a justification of the SCBA as the large N limit of Weyl fermions. 

\end{abstract}

\maketitle

\section{Introduction}

A very common and straightforward approach to the average one-particle Green's function of a disordered 
system of non-interacting electrons is the self-consistent Born approximation (SCBA). 
Numerous applications of SCBA-based techniques to low--dimensional systems of disordered electrons provided an excellent 
confirmation for a number of experimental 
observations~[\onlinecite{Fradkin1986,Lee1993,Shon1998,Ando2002,Zheng2002}]. 
However, the claim of a 'failure of the SCBA' for two-dimensional (2d) Dirac particles by 
Aleiner and Efetov~[\onlinecite{Aleiner2006}] has questioned whether this approach is applicable at all to two-band systems with
spectral degeneracies in general. These authors argued that the self-energy diagrams of order $g^2$ in disorder 
strength dominate over the local density of states (DOS) predicted by the SCBA. Another approach to the average
DOS, based on a bosonization concept, predicts that the DOS obeys a power law at the node with a disorder
dependent exponent~[\onlinecite{Tsvelik1994}], which also contradicts the non-vanishing DOS of the
SCBA. 

%All this was 
The SCBA and the power-law prediction for the average DOS were
checked recently in numerical studies as well as by a functional renormalization group approach 
by Sbierski et al., who found that there is no power law and that the SCBA is only missing a 
factor of 2 in the logarithm of the DOS (i.e., the square root of SCBA must be taken)~[\onlinecite{Sbirski2017}]
in 2d. Moreover, in 3d the critical disorder strength is twice as large for the SCBA, although the slope of the
DOS agrees quite well~[\onlinecite{Sbirski2014,Sbirski2017}]. 
These missing factors of 2 suggest that the corrections to the SCBA are of the same order
as the SCBA itself. The renewed interest in the behavior of the average DOS of Weyl 
systems~[\onlinecite{Sbirski2014,Sbirski2017,pixley15,pixley17,ziegler17}] suggests a clarification of the
role of the traditional SCBA approach, which is based on a commonly accepted mean-field type of approximation. 
Since the SCBA is equivalent to a saddle-point approximation of a
functional integral, these corrections are easily accessible from the fluctuations around the
saddle point. The aim of the present work is to analyze corrections to the 
SCBA for 2d and 3d disordered Weyl fermions in a systematic $1/N$ expansion~[\onlinecite{
Ye1998,Ye1999,Herbut2006,Son2007,Forster2008}].

\section{The model}

\no
%The aim of this paper is to shed light on the structure of the DOS in a system of
%non-interacting two and three dimensional Weyl (or Dirac) electrons in a random environment. 
A generalization of the Weyl Hamiltonian to the one with $N$ orbitals per site 
reads~[\onlinecite{Oppermann1979,Wegner1979,ziegler97a}]
\begin{equation}
\label{eq:Weyl}
H =  {\bf 1}^{}_N\otimes i\slashed\partial + v\otimes\sigma^{}_{\nu}
\ , 
\end{equation}
where the Dirac contraction notation is  
$\slashed\partial = \sigma^{}_1\partial^{}_1 + \sigma^{}_2\partial^{}_2$ in 2d and 
$\slashed\partial = \sigma^{}_1\partial^{}_1 + \sigma^{}_2\partial^{}_2 + \sigma^{}_3\partial^{}_3$
in 3d, respectively. In contrast to the 2d case, in 3d there is no Pauli matrix left which anticommutes with the Dirac operator. 
The Pauli matrix $\sigma^{}_{\nu}$ is either $\sigma^{}_0$ (i.e., the $2\times2$ unit matrix) for a random scalar potential, 
or $\sigma^{}_3$ for a random Dirac mass in 2d, and only $\sigma^{}_0$ for a random scalar potential in 3d.  
Different physical realizations of Weyl electrons reveal different values of  $N$, 
e.g., $N=2$ for graphene, $N=4$ for $\pi$-flux model in 2d, and $N=8$ for $\pi$-flux model in 
3d~[\onlinecite{pixley15,pixley17}]. 
The random potential $v$ represents a symmetric $N\times N$-matrix $(v_r^{ij})_{i,j=1,...,N}$
with zero mean $\langle v_r^{ij}\rangle =0$ and with the correlator 
\begin{equation}
\label{N_corr}
\langle v^{ij}_{r} v^{kl}_{r^\prime} \rangle^{} =  \frac{g}{N}\delta^{}_{il}\delta^{}_{jk}\delta(r-r^\prime)
\ .
\end{equation}
We use the convention
$\hbar v^{}_F=1$, where $v_F$ denotes the Fermi velocity. 

The DOS is the imaginary part of the diagonal element of the retarded Green's function
$G^{}(i\epsilon)=[{\bf 1}^{}_N\otimes G^{-1}_0  + v\otimes\sigma^{}_{\nu}]^{-1}$ ($\epsilon>0$), where 
$G_0 =[i\epsilon\sigma^{}_0 + i\slashed\partial]^{-1}$ is the one-particle Green's function of the electron in a clean system:
%,summed over all copies $i=1,...,N$
\begin{equation}
\label{eq:DOS} 
\varrho(\epsilon) = -\frac{1}{\pi}{\rm Im}~{\rm Tr}_{2N}~G^{}_{rr}(i\epsilon)
\ .
\end{equation}
The operator ${\rm Tr}_{2N}$ denotes the trace with respect to the space of Pauli matrices and
the $N$ orbitals.
%and the summation over all Weyl copies $i$ is understood. 
The Green's function can be written as a functional integral 
\begin{equation}
\label{eq:SuSy}
 G^{ii}_{rr} = -i\int{\cal D}\psi^\dag{\cal D}\psi{\cal D}\varphi^\dag{\cal D}\varphi~\psi^{i}_{r}\psi^{i\dag}_{r}~
 e^{i{\cal S}^{}_F+i{\cal S}^{}_B}
 \ ,
\end{equation}
with the actions
\begin{equation}
\label{eq:SB}
{\cal S}^{}_B  = \varphi^\dag\cdot[{\bf 1}^{}_N\otimes G^{-1}_0  + v\otimes\sigma^{}_{\nu}]\varphi^{} ,
\end{equation}
and 
\begin{equation}
\label{eq:SF}
{\cal S}^{}_F = 
\psi^\dag\cdot[{\bf 1}^{}_N\otimes G^{-1}_0  + v\otimes\sigma^{}_{\nu}]\psi^{}
\ .
\end{equation}
Here,  $\varphi^{}$ represents a $2N$-component complex field and $\psi^{}$ a $2N$-component
Grassmann field. Since $\epsilon>0$, the convergence of the complex functional integral is guaranteed. The advantage of using complex and Grassmann fields is that the integral 
\begin{equation}
\int {\cal D}\psi^\dag{\cal D}\psi{\cal D}\varphi^\dag{\cal D}\varphi~e^{i{\cal S}^{}_F+i{\cal S}^{}_B}=1
\end{equation}
is already normalized, whereas using only the complex or only the Grassmann part requires an extra normalization. This would 
create problems for the calculation of the average with respect to disorder.

Arranging bosonic and fermionic fields to a vector superfield $\Phi^{}=(\varphi,\psi)^{\bf T}$ 
we can easily perform the disorder averaging (cf. Appendix~\ref{app:HST}), and decouple by means
of the matrix superfield $\hat Q$. This has the matrix structure
\begin{equation}
\label{eq:HatField}
\hat Q_r = \left(
\begin{array}{cc}
Q_r & \chi_r \\
\bar\chi_r & iP_r
\end{array}
\right)
\ , 
\end{equation}
with $Q_r,P_r$ representing $2\times 2$ matrices with commuting and $\chi_r,\bar\chi_r$ with anticommuting 
matrix elements. The average Green's function then becomes %in functional integral representation 
\begin{equation}
\label{eq:GF1}
\sum_{i=1}^N\langle G^{ii}_{rr}(i\epsilon)\rangle = -i\frac{N}{g}\sigma^{}_{\nu}\int{\cal D}\hat Q~P_r~e^{-N{\cal S}[\hat Q]}
\ ,
\end{equation}
with the effective action
\begin{equation}
{\cal S}[\hat Q] = {\rm trg}\left\{\frac{1}{2g}\hat Q^2 +  \log[\sigma^{}_0\otimes G^{-1}_0 + \hat Q\Sigma^{}_{\nu}]\right\}
\ ,
\end{equation}
where $\Sigma^{}_{\nu}=\sigma^{}_0\otimes\sigma^{}_{\nu}$, and ${\rm trg}$ is the graded trace.
Since the effective action does not depend on $N$, the integral in Eq. (\ref{eq:GF1}) suggests a saddle-point
approximation for large $N$ and a $1/N$-expansion. This will be discussed subsequently.

The scattering rate $\eta$ is related to the nontrivial saddle point of 
${\cal S}[\hat Q]$, which is a solution of the saddle-point equation $\delta{\cal S}=0$. %$\partial_{\hat Q}{\cal S}=0$.
Thus, the field can be written as ${\hat Q}={\hat Q}_0+{\hat Q}'$ with the saddle point
\begin{equation}
{\hat Q}_0 = i\eta\Sigma^{}_\nu
%\left(
%\begin{array}{cc}
%i\eta \Sigma^{}_{\nu} & 0 \\
%0 & -\eta \Sigma^{}_{\nu}
%\end{array}
%\right)
\ .
\end{equation}
For convenience, we rename the integration field ${\hat Q}'\to {\hat Q}$.
In terms of the Green's function we get
\begin{equation}
\label{eq:GF2}
\sum_{i=1}^N\langle G^{ii}_{rr}\rangle = -iN\frac{\eta}{g}\sigma^{}_0 + \sum_{i=1}^N\delta G^{ii}_{rr}
\ ,
\end{equation}
where the first term represents the uniform saddle-point contribution through the scattering rate 
and the second term represents the correction due to quantum fluctuations around the saddle point
\begin{equation}
\sum_{i=1}^N\delta G^{ii}_{rr} = -i\frac{N}{g}\sigma^{}_{\nu}\int{\cal D}\hat Q_r~P_r~
e^{-N{\cal S}[\hat Q]}
\ ,
\label{funct_int2}
\end{equation}
with the shifted action
\begin{equation}
\label{eq:ShiftAct}
{\cal S}[\hat Q] = {\rm trg}\left\{\frac{1}{2g}(\hat Q+i\eta\Sigma^{}_{\nu})^2 
+ \log[{\bar G}^{-1} + \hat Q\Sigma^{}_{\nu}]\right\} .
\end{equation}
The inverse average Green's function reads 
${\bar G}^{-1} = \sigma^{}_0\otimes[iz\sigma^{}_0 + i\slashed\partial]$, $z=\epsilon+\eta$. 
From Eqs.~(\ref{eq:DOS}) and (\ref{eq:GF2}) the saddle-point approximation of the DOS becomes 
\begin{equation}
\label{eq:DOSmf}
\varrho^{}_{SCBA} = \frac{2N}{\pi}\frac{\eta}{g}
\ ,
\end{equation}
regardless of the model dimension $d$. The behavior of the scattering 
rate $\eta$ does however crucially depend on $d$.

\section{Saddle-point analysis and fluctuations}
 
In order to obtain the effective action in the limit of slowly varying quantum fields we expand
Eq.~(\ref{eq:ShiftAct}) in powers of fluctuations $\hat Q$ around the saddle point.
The small expansion parameter is $1/N$, 
since only the prefactor of the action depends on $N$ in Eq. (\ref{funct_int2}).
Therefore, we can employ a saddle-point approximation, which
%in the limit $\epsilon\to0$ this 
leads to the saddle-point condition
\begin{equation}
\label{eq:SPC}
\eta
%= -\langle r|[iz +i\slashed\partial]^{-1}|r\rangle 
= {g}\int\frac{d^dq}{(2\pi)^d}~\frac{\eta}{\eta^2+q^2}
\ .
\end{equation}
Solutions of this equation are described in the literature~[\onlinecite{Fradkin1986,Lee1993,Ziegler1997,Ziegler2015}]. 
While in 2d they predict an exponentially small but non-vanishing scattering rate for any value of the 
disorder strength $g$
\begin{equation}
\eta^{}_{2d} \sim \Lambda e^{-2\pi/g}
\ ,
\end{equation}
in 3d the non-vanishing scattering rate emerges only if the disorder strength becomes larger than a critical value 
\begin{equation}
g^{}_c \sim \frac{2\pi^2}{\Lambda}
\ ,
\end{equation}
where $\Lambda$ represents a UV-cutoff of the order of inverse lattice constant, giving for small values of $g$ %above the critical 
\begin{equation}
\eta^{}_{3d} \sim g\left(\frac{2\pi}{g^{}_c}\right)^2\theta(g-g^{}_c)
\ .
\end{equation}
The expansion of the logarithm around this nontrivial vacuum reads
\begin{equation}
{\cal S}[\hat Q] 
={\rm trg}\left\{
\frac{1}{2g}{\hat Q}^2 - \frac{1}{2} [{\bar G}\hat Q\Sigma^{}_{\nu}]^2
\label{eq:Exp}
-\sum_{n\geqslant3}\frac{(-1)^n}{n}[{\bar G}\hat Q \Sigma^{}_{\nu}]^n
\right\}.
\end{equation}
The third term represents a perturbation to the scattering rate beyond the Gaussian approximation. 
Because of the structure of the matrix ${\bar G}$ all sectors of our theory, bosonic and fermionic, have 
the same propagators;
i.e., there is no supersymmetry breaking. 
%To Gaussian order, we keep track on the subspace 
%$P$, since the  field monomial under the functional integral contains only $P$'s (below we restore for a %while the original scaling of the fields $\hat Q$): 
In Gaussian order of $P_r$ we get
\begin{eqnarray}
 {\cal S}^{}_G[P] = {\rm tr}\left\{\frac{1}{2g} P^2 - \frac{1}{2} {\bar G}P\sigma^{}_{\nu} {\bar G}P\sigma^{}_{\nu}\right\}
\label{action3}
\end{eqnarray}
with the Hermitean matrix field $P$, which can be represented as
\begin{eqnarray}
P &=& \left(
\begin{array}{ccc}
 P_0 + P_3   & P_1 - i P_2 \\
 P_1 + i P_2 & P_0 - P_3
\end{array}
\right)= P^{}_\alpha\sigma^{}_\alpha %+ P_1\sigma_1 + P_2\sigma_2 + P_3\sigma_3
\end{eqnarray}
with real $P^{}_\alpha$. The summation convention is used in Eq. (\ref{action3}) for
$\alpha = 0,1,2,3$, and $\sigma^{}_\alpha$ denote the Pauli matrices. 
This enables us to perform the trace in the first term immediately:
\begin{equation}
{\rm tr}\frac{1}{2g}P^2  = \frac{1}{g}P\cdot P = \frac{1}{g}\int\frac{d^dq}{(2\pi)^d} P^{}_q\cdot P^{}_{-q}
\end{equation}
where the vector $P$ is assembled from elements of the matrix $P^{}_\alpha$. 
Second term reads after transforming it into Fourier representation 
\begin{equation}
\label{eq:Vert}
\frac{1}{2}{\rm tr}{\bar G}P\sigma^{}_{\nu} {\bar G}P\sigma^{}_{\nu} 
%&=& \frac{1}{2}{\rm Tr}\int\frac{d^2q}{(2\pi)^2} \int\frac{d^2p}{(2\pi)^2}~G(p)P_q\sigma^{}_{\nu} {\bar G}(q-p)P_{-q}\sigma^{}_{\nu} \\ 
= \int\frac{d^dq}{(2\pi)^d}~ P^\alpha_q P^\beta_{-q} \Gamma^{(\nu)}_{2|\alpha\beta}(q),
\end{equation}
where $\nu=0$ denotes the random scalar potential and $\nu=3$ a random gap. 
The explicit expression and evaluation of the two-point vertex function $\Gamma^{(\nu)}_{2}$ are 
given in Appendix~\ref{app:EffProp}. It turns out that the inverse effective propagators do not 
have zero modes. This reflects the absence of a broken continuous symmetry.
For vanishing momenta and frequencies, the effective action becomes
\begin{equation}
\label{eq:LocAc}
{\cal S}^{}_G[\hat Q] = {\rm M}^{(\nu)}_{aa}\int d^dr ~ \left[Q^a_rQ^a_r +2 \chi^a_r\bar\chi^a_r +P^a_rP^a_r\right]
\ .
\end{equation}
In 2d the solution $\eta=0$ of Eq.~(\ref{eq:SPC}) is always unstable (cf. Appendix~\ref{app:EffProp}), and for $\eta>0$
the mass matrices are 
\begin{equation}
\label{eq:Mass02d}
{\rm M}^{(0)}_{2d} \sim \left(
\begin{array}{cccc}
 \frac{c_\Lambda}{2\pi} & 0 & 0 & 0 \\
0 &  \frac{1}{g} + \frac{c_\Lambda}{2\pi} & 0 & 0 \\
0 & 0 &  \frac{1}{g} + \frac{c_\Lambda}{2\pi} & 0 \\
0 & 0 & 0 & \frac{2}{g}
\end{array}
\right)\ ,\ c_\Lambda=\frac{\Lambda^2}{\eta^2+\Lambda^2}
\end{equation}
for a random scalar and
\begin{equation}
\label{eq:Mass32d}
{\rm M}^{(3)}_{2d} \sim \left(
\begin{array}{cccc}
\frac{2}{g}  & 0 & 0 & 0 \\
0 &  \frac{1}{g} - \frac{c_\Lambda}{4\pi} & 0 & 0 \\
0 & 0 &  \frac{1}{g} - \frac{c_\Lambda}{4\pi} & 0 \\
0 & 0 & 0 & \frac{c_\Lambda}{2\pi}
\end{array}
\right)
\ ,
\end{equation}
for a random gap. 
In the 3d case, the mass matrix for the random potential reads
\begin{equation}
\label{eq:Mass03d}
{\rm M}^{(0)}_{3d} \sim \left(
\begin{array}{cccc}
\frac{\eta}{4\pi} & 0 & 0 & 0 \\
0 & {\rm\lambda}  & 0 & 0 \\
0 & 0 & {\rm\lambda} & 0 \\
0 & 0 & 0 & {\rm\lambda}
\end{array}
\right),
\end{equation}
with the matrix element ${\rm\lambda}$ given in Appendix~\ref{app:EffProp}, Eq.~(\ref{eq:MassEls3d}).

\subsection{Corrections to the DOS: weak disorder} 

\begin{figure}[t]
\includegraphics[width=3.0cm]{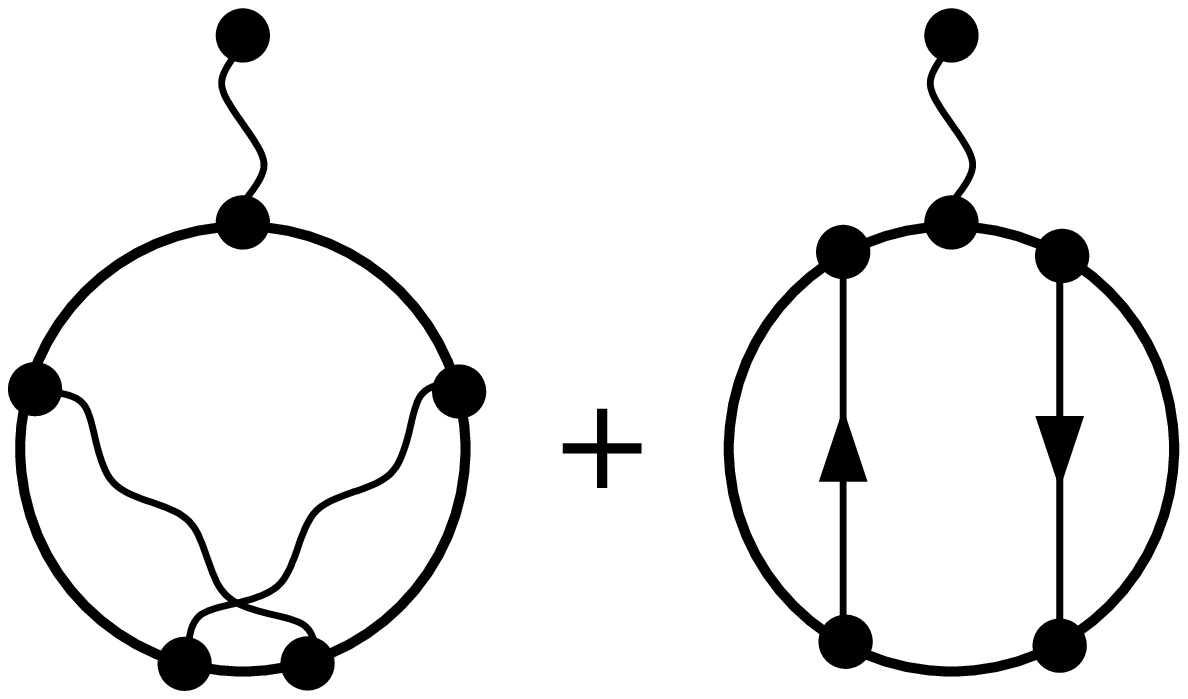}
\caption{Leading corrections of the DOS to the SCBA in the $1/N$ expansion.}
\label{fig:5Ordc}
\end{figure}

The calculation of the DOS corrections can be organized in terms of a $1/N$ expansion, which is obtained by
rescaling the field ${\hat Q}$ with $\sqrt{N}$. This absorbs the prefactor $N$
in the exponent of Eq. (\ref{funct_int2}) into the quadratic order of the expansion
in Eq. (\ref{eq:Exp}) and produces powers of $1/\sqrt{N}$ for higher order terms. 
A further simplification comes from the assumption of weak disorder; i.e., $g\ll\Lambda^2$. 
In this case only the smallest diagonal element of the mass matrices in Eqs. 
(\ref{eq:Mass02d})--(\ref{eq:Mass03d}) dominates the Gaussian fluctuations.
Taking the 
momentum dependence to leading order in a gradient expansion into account, 
the corresponding excitation mode becomes
\begin{equation}
\label{eq:DisProp}
\Pi^{(\nu)}_{\nu\nu}(q) \sim \frac{12\pi}{g} \frac{\eta^{4-d}}{q^2 + m^2_d}, 
\end{equation}
where the masses are $m^2_2 = 12\eta^2$ in 2d and $m^{2}_3=6\eta^2$ in 3d
(cf. Appendix~\ref{app:EffProp}). 
Since the supersymmetry remains unbroken, the correlation function of the Grassmann field 
$\bar\chi\chi$ and the Hermitean field $Q$  are given by the same expression. 
In position space the correlator Eq.~(\ref{eq:DisProp}) decays exponentially (in 2d it is 
proportional to the modified Bessel function of second kind $K^{}_0[m^{}_2|r-r^\prime|]$, 
in 3d to $\exp[-m^{}_3|r-r^\prime|]/|r-r^\prime|$) and can be crudely approximated by the 
Dirac delta function
\begin{equation}
\label{eq:DisPropPos}
\Pi^{(\nu)}_{\nu\nu}(r,r^\prime) \sim  \delta(r-r^\prime)
\ .
\end{equation}
%which enables us to capture the essential physics and is particularly convenient for practical %calculations. 
Then the first non-vanishing correction to the average one-particle Green's function 
(cf. Fig. \ref{fig:5Ordc}) is of the order $1/N^2$ and reads
\begin{equation}
\label{eq:corr_2d}
\varrho = \varrho^{}_{SCBA}  + \frac{\varrho^{}_{SCBA}}{N^2}\left(\frac{2}{3}-\frac{g}{2\pi}-\frac{g^3}{4\pi^3}\right)
+O(N^{-2})
\end{equation}
in 2d and
\begin{eqnarray}
\nn
\varrho &=& \varrho^{}_{SCBA}  + \frac{\varrho^{}_{SCBA}}{N^2}
\left[\frac{\pi^2}{2}\left(1 - \frac{5\pi}{19}\right)  \right.
\\
\label{eq:corr_3d}
&&- 
\left.\frac{g\eta}{2\pi} - \frac{g^2\eta^2}{8}\right] +O(N^{-2})
\end{eqnarray}
in 3d for the DOS of Eq. (\ref{eq:DOS}). Details of the $1/N$ expansion
are presented in Apps. \ref{app:perturbation} and \ref{app:PTloop}.

Our calculation identifies the 'sunrise' (or maximally crossed) diagrams as dominant. 
This gives corrections in each order $1/N$ with a polynomial in $g$. In other words,
for each order $1/N$ there are $g$-independent contributions to DOS.
Since, however, an $n$-order correction goes proportionally to $N^{1-n}$, the series converges rapidly.

\subsection{Corrections to the DOS: strong disorder} 

In the regime of large $g$ values we can neglect the gradient terms in the correlators.  
In 2d and for scalar disorder the correlator of the fields $P^{}_0$ then reads 
\begin{equation}
\langle P^{}_{0r}P^{}_{0r^\prime} \rangle \sim \frac{\pi}{g}\delta(r-r^\prime),
\end{equation}
which is needed in order for DOS to have a finite trace, while that of $P^{}_{i=1,2}$
\begin{equation}
\langle P^{}_{ir}P^{}_{ir^\prime} \rangle \sim \frac{\pi}{2\pi + gc^{}_\Lambda}\delta(r-r^\prime),
\end{equation}
which is negligible in comparison to the correlator of the fields $P^{}_3$
\begin{equation}
\langle P^{}_{3r}P^{}_{3r^\prime} \rangle \sim \frac{1}{4}\delta(r-r^\prime).
\end{equation}
For the case of the random mass disorder the situation is analogous, with the interchanging role of the fields $P^{}_0$ and $P^{}_3$. 
The dominant contribution comes from the diagrams with the loops which couple to the external field via $P^{}_0$ channel 
and internal coupling of $P^{}_3$ fields. In Appendix~\ref{app:PTloopSt} we obtain for 2d
\begin{equation}
\label{eq:corr_2dstrong}
\varrho = \varrho^{}_{SCBA}  + \frac{\varrho^{}_{SCBA}}{(4N)^2}\left(\frac{3}{2} - \frac{2\pi}{g} \right)
+O(N^{-2}),
\end{equation}
i.e. the correction is positive. In 3d, the correlator of $P^{}_0$ in strong disorder limit reads
\begin{equation}
\langle P^{}_{0r}P^{}_{0r^\prime} \rangle \sim \frac{2\pi}{g\eta}\delta(r-r^\prime), 
\end{equation}
while that of $P^{}_{i=1,2,3}$ is
\begin{equation}
\langle P^{}_{ir}P^{}_{ir^\prime} \rangle \sim  \frac{1}{g\Lambda}\delta(r-r^\prime),
\end{equation}
cf. Eq.~(\ref{eq:MassEls3d}) and since $\Lambda\gg\eta$ they are parametrically smaller and 
can be neglected in crudest approximation. We get 
\begin{eqnarray}
\nn
\varrho &=& \varrho^{}_{SCBA}  - \frac{\varrho^{}_{SCBA}}{N^2}
\left[  \frac{\pi^3}{g\eta} + \frac{(2\pi)^2}{g^2\eta^2} \right.
\\
\label{eq:corr_3dstrong}
&&\left. 
-\frac{(2\pi)^5}{8g^3\eta^3}\left(1 - \frac{5\pi}{19}\right)
\right] +O(N^{-2}),
\end{eqnarray}
i.e. the corrections are negative for large $g$.

\section{Discussion}

%Recently the behavior of the DOS of weakly disordered Weyl electron gas gained a big deal of attention.
Our analysis of the perturbative expansion for the average DOS in terms of disorder strength $g$ clearly
indicates that this expansion cannot be organized in a systematic way in powers of $g$. This result seems
to support the claim of Aleiner and Efetov~[\onlinecite{Aleiner2006}] of a ``failure of the SCBA''. However, 
using $N$ copies of Weyl fermions, as described by the model in Eqs. (\ref{eq:Weyl}), (\ref{N_corr}),
provides a systematic $1/N$ expansion which reveals that the SCBA has only 
corrections of order $1/N$. Moreover, the corrections up to order $1/N$ in Eqs. (\ref{eq:corr_2d}), (\ref{eq:corr_3d})
(\ref{eq:corr_2dstrong}), and (\ref{eq:corr_3dstrong})
give an enhancement of the DOS in comparison to the SCBA, in agreement with the numerical 
results found by Sbirski et al.~[\onlinecite{Sbirski2014,Sbirski2017}] who got a doubling of the SCBA values in 2d~[\onlinecite{Sbirski2017}]. 
The $1/N$ expansion suggests that this doubling is specific for a single-component Weyl fermion.
For models with larger $N$, e.g., for different versions of the $\pi$-flux model~[\onlinecite{pixley15,pixley17}], the DOS 
corrections to the SCBA become virtually negligible and the SCBA is exact in the limit $N\to\infty$. 
Finally, we don't find a shift of the critical disorder strength $g_c$ in the $1/N$ expansion
for the appearance of a nonzero DOS in 3d Weyl fermions, which was predicted in Refs.~[\onlinecite{Sbirski2014,Sbirski2017}].

The behavior of the leading order correction as a function of the disorder strength depends 
crucially on the spatial dimension of the system. For instance, in zero dimension (random matrix model) 
the saddle point condition reads $g=\eta^2$ which results in a $g$-independent DOS.

\section{Conclusions}

Our extended Weyl-fermion model with $N$ orbitals per site gives in the $N\to\infty$ limit
for the average DOS the SCBA result and a systematic $1/N$ expansion for the corrections to the SCBA
at finite $N$. Each term in the $1/N$ expansion depends on the disorder parameter $g$, which can be
expanded for weak disorder as a power series of $g$ or for strong disorder as a power series
of $1/g$. This result demonstrates the reliability of the SCBA and the existence of a systematic
expansion for disordered Weyl fermions at the node.

\section*{ACKNOWLEDGMENTS}

This work was supported by a grant of the Julian Schwinger Foundation for Physical Research.

\vskip0.5cm

\appendix

\section{Hubbard-Stratonovich transformation}
\label{app:HST}

The graded trace of a matrix 
\begin{eqnarray}
\label{eq:trg}
M=\left( 
\begin{array}{cc}
 A & B \\
 C & D
\end{array}
\right) 
\end{eqnarray}
with quadratic matrices $A,B,C,D$ reads 
\begin{eqnarray}
{\rm Trg}~M = {\rm Tr}[A-D]. 
\end{eqnarray}
The graded determinant of the matrix $M$ reads
\begin{eqnarray}
{\rm detg}~M = \frac{\det A}{\det D}\det[1-BD^{-1}CA^{-1}].
\end{eqnarray}

%\subsection{Averaging procedure}
The ensemble average of the Green's function 
reads
\begin{eqnarray}
\langle G^{}_{rr}\rangle &=& {\cal N}^{-1}\int^{\infty}_{-\infty}dv~{\cal P}(v)~G^{}_{rr},\\
{\cal N} &=& \int^{\infty}_{-\infty}dv~{\cal P}(v),\; {\cal P}(v) = e^{-\frac{N}{2g}{\rm tr}v^2},
\end{eqnarray}
where the integrals are the functional ones. Consider the combination of $v$-dependent terms in the exponent of the integral:
\begin{eqnarray}
\nn
&&
\frac{N}{2g}{\rm tr}v^2 - i\Phi^\dag\cdot[{\bf 1}^{}_2\otimes v\otimes\sigma^{}_\nu]\Phi \\ 
\nn
&&
= \frac{N}{2g}v^2_{ab} - i v^{}_{ab}[\varphi^\dag_a\sigma^{}_\nu\varphi^{}_b + \psi^\dag_a\sigma^{}_\nu\psi^{}_b]\\
\nn
&& 
= \frac{N}{2g}v^2_{ab} - i v^{}_{ab}~\Phi^\dag_a\Sigma^{}_\nu\Phi^{}_b ,
\end{eqnarray}
where $\Sigma^{}_{\nu}=\sigma^{}_0\otimes\sigma^{}_{\nu}$. Here, the summation 
convention is understood. The integration over $v^{}_{ab}$ can be performed after 
completing the square and shifting the potential matrix elements as
\begin{equation}
v^{}_{ab} \to v^{}_{ab} - i \frac{g}{N}\Phi^\dag_a\Sigma^{}_\nu\Phi^{}_b. 
\end{equation}
What remains is the interaction term
\begin{equation}
-\frac{g}{2N}{\rm trg}\left(\Sigma^{}_\nu\Phi^{}_a\Phi^\dag_a\Sigma^{}_\nu\Phi^{}_b\Phi^\dag_b \right),
\end{equation}
which gives the full action 
\begin{eqnarray}
\nn
\bar{\cal S}^{}_{BF}[\Phi^\dag,\Phi] &=& i\Phi^\dag\cdot {\bf 1}^{}_N\otimes G^{-1}_0\Phi^{} \\
&-& 
\frac{g}{2N} {\rm trg}[\Sigma^{}_{\nu}\Phi^{}_a\Phi^\dag_a]^2.
\end{eqnarray}
The interaction term is then decoupled by means of a Hubbard-Stratonovich transformation
\begin{eqnarray}
\nn
&\displaystyle 
\bar{\cal S}^{}_{BF}[Q,\Phi^\dag,\Phi] = \frac{N}{2g}{\rm trg}[\hat Q-i\frac{g}{N}\Sigma^{}_{\nu}\Phi^{}_a\Phi^\dag_a]^2 .&\\
\label{eq:DecAct}
&\displaystyle
- i\Phi^\dag\cdot{\bf 1}^{}_N\otimes G^{-1}_0\Phi^{} + \frac{g}{2N} {\rm trg}[\Sigma^{}_{\nu}\Phi^{}_a\Phi^\dag_a]^2,&\;\;
\end{eqnarray}
where we shifted the matrix field $\hat Q$ exploiting the "translational invariance" of the corresponding functional integral measure. 
The element $\sim (\Sigma^{}_\nu\Phi^{}_{a}\Phi^\dag_a)^2$ vanishes, but the form of 
Eq.~(\ref{eq:DecAct}) is useful in order to recognize the structure of the integrand. 
For every copy it can be expressed in terms of the matrix field $\hat Q$ as follows:
\begin{equation}
\label{eq:MatElt}
\psi^{}_r\psi^\dag_r =
i\frac{1}{g}\sigma^{}_{\nu}[iP - i\frac{g}{N}\sigma^{}_{\nu}\Phi^{}_{2a}\Phi^\dag_{2a} - iP]^{}_r.
\end{equation}
Inserting this expression into the functional integral we notice that the integration over the term $P - (g/N)\sigma^{}_{\nu}\Phi^{}_{2a}\Phi^\dag_{2a}$ can be performed independently and in the position space, since the term with $\hat Q-i(g/N)\Sigma^{}_{\nu}\Phi^{}_a\Phi^\dag_a$ does not possess any gradients and therefore is already diagonal. The contribution from this term is zero. Then combining Eq.~(\ref{eq:DecAct}) and Eq.~(\ref{eq:MatElt}) we get
\begin{equation}
\langle G^{ii}_{rr}\rangle = 
-\frac{i}{g}\sigma^{}_{\nu}\int{\cal D}\hat Q~P^{}_r~\int{\cal D}\Phi^\dag{\cal D}\Phi ~ e^{-\bar{\cal S}^{}_{BF}[\hat Q,\Phi^\dag,\Phi]},
\end{equation}
at which point the integration over vector fields can be carried out. Rising the graded determinant into the exponent we acquire the log-term in Eq.~(\ref{eq:GF1}).

\begin{widetext}
\section{Effective propagators and correlation functions}
\label{app:EffProp}

Below we always send the UV-cutoff of radial integrals to infinity if the dimensional analysis points out their convergence. In the infrared, the divergences are cut off by the scattering rate $\eta$. The two-point vertex functions which appear in Eq.~(\ref{eq:Vert}) reads 
\begin{equation}
 \Gamma^{(\nu)}_{2|\alpha\beta}(q) = \frac{1}{2}{\rm Tr} \int\frac{d^dp}{(2\pi)^d}~ \frac{\displaystyle [-iz+\slashed p]\sigma^{}_\alpha\sigma^{}_{\nu}[-iz+\slashed q+\slashed p]\sigma^{}_\beta\sigma^{}_{\nu}}{[z^2 + p^2][z^2 + (p+q)^2]}.
\end{equation}
In order to calculate the contributions to the mass we set $\epsilon=0$ and $p=0$. We first neglect all terms under the integral which are not rotationally invariant:
\begin{eqnarray}
\label{eq:Gamma2b}
\Gamma^{(\nu)}_{2|\alpha\beta} = \frac{1}{2}{\rm Tr}\int\frac{d^dp}{(2\pi)^d}~
\frac{\displaystyle -\eta^2\sigma^{}_\alpha\sigma^{}_{\nu}\sigma^{}_\beta\sigma^{}_{\nu} +\frac{p^2}{d} \sigma^{}_{i=1,\dots,d}\sigma^{}_{\alpha=0,\dots,3}\sigma^{}_{\nu}\sigma^{}_i\sigma^{}_\beta\sigma^{}_{\nu}}{[\eta^2+p^2]^2},
\end{eqnarray}
where the factor $1/d$ in front of the second part appears due to the angular average. Since the product of any two or three Pauli matrices is a Pauli matrix again, the trace in Eq.~(\ref{eq:Gamma2b}) is non-zero only for $\alpha=\beta$. Therefore the inverse propagator is diagonal in both 2d and 3d. Below we evaluate  Eq.~(\ref{eq:Gamma2b}) for all combinations of external indices $\alpha,\beta$ and use following short-hands
$\displaystyle 
\sigma^{}_i \sigma^{}_{\nu} = \zeta \sigma^{}_{\nu}\sigma^{}_i$ for all $i$. In 2d $\zeta = (-)+1,$ if $\sigma^{}_{\nu}$ (anti)commutes  with 
$\sigma^{}_i$, in 3d $\zeta=+1$. The trace of the term proportional to  $\eta^2$  gives $\displaystyle{\rm Tr}~\sigma^{}_\alpha\sigma^{}_{\nu}\sigma^{}_\beta\sigma^{}_{\nu} = 2\zeta \delta^{}_{\alpha\beta}$. Second part has to be evaluated for different index combinations separately. In 2d: 1) $\alpha=\beta=1,2$ is zero because of the matrix product property $\sigma^{}_i\sigma^{}_a\sigma^{}_i = \sigma^{}_a(\sigma^{}_{i=\alpha}-\sigma^{}_{i\neq\alpha})\sigma^{}_i = 0$, $i$ is summed over; 
2) $\alpha=\beta=0$:  
$\displaystyle 
{\rm Tr}[-\eta^2 \sigma^{}_{\nu}\sigma^{}_{\nu} + p^2/2 \sigma^{}_i\sigma^{}_{\nu}\sigma^{}_i\sigma^{}_{\nu}] = 2(\zeta p^2 -\eta^2); 
$
3) $\alpha=\beta=3$, i.e. $\sigma^{}_{\nu}$ commutes with $\sigma^{}_3$ for both disorder types:
$\displaystyle 
{\rm Tr}~[ -\eta^2 \sigma^{}_3\sigma^{}_{\nu}\sigma^{}_3\sigma^{}_{\nu} + p^2/2 \sigma^{}_i\sigma^{}_3\sigma^{}_{\nu}\sigma^{}_i\sigma^{}_3\sigma^{}_{\nu}] = -2(\eta^2 + \zeta p^2)$.
With help of the the saddle-point condition Eq.~(\ref{eq:SPC}), cf. Ref~[\onlinecite{Ziegler1997}], the elements of the mass matrix become 
\begin{eqnarray}
{\rm M}^{(\nu)}_{00} &=& \frac{1}{g} - \int\frac{d^2p}{(2\pi)^2}~\frac{-\eta^2+\zeta p^2}{[p^2+\eta^2]^2} =
\left\{
\begin{array}{cc}
\displaystyle \frac{\Lambda^2}{2\pi(\eta^2+\Lambda^2)}, & \nu=0, \,\,\zeta=+1 \\ 
\\
\displaystyle \frac{2}{g}, & \nu=3,\;\;\zeta=-1
\end{array}
\right. ,
\\\nn\\
{\rm M}^{(\nu)}_{\alpha\alpha=1,2} &=& \frac{1}{g} + \int\frac{d^2p}{(2\pi)^2}~\frac{\zeta\eta^2}{[p^2+\eta^2]^2} = 
\frac{1}{g} + \frac{\zeta}{4\pi} = 
\left\{
\begin{array}{cc}
\displaystyle \frac{1}{g} + , & \nu=0, \,\,\zeta=+1 \\ 
\\
\displaystyle \frac{1}{g} - \frac{\Lambda^2}{4\pi(\eta^2+\Lambda^2)}, & \nu=3,\;\; \zeta=-1
\end{array}
\right.,
\\\nn\\
{\rm M}^{(\nu)}_{33} &=& \frac{1}{g} - \int\frac{d^2p}{(2\pi)^2}~\frac{-\eta^2 - \zeta p^2}{[p^2+\eta^2]^2} = 
\left\{
\begin{array}{cc}
\displaystyle \frac{2}{g}, & \nu=0, \,\,\zeta=+1 \\ 
\\
\displaystyle \frac{\Lambda^2}{2\pi(\eta^2+\Lambda^2)}, & \nu=3,\;\;\zeta=-1
\end{array}
\right. .
\end{eqnarray}

In 3d the evaluation differs technically in that respect, that there is no Pauli matrix 
which anticommutes with the kinetic energy operator $-i\slashed\partial$. Second term is for 1)  
$\sigma^{}_\alpha \neq 0,\;\;\sigma^{}_\beta \neq 0$:
$\displaystyle 
{\rm Tr}~\sigma^{}_\alpha\sigma^{}_i\sigma^{}_\beta\sigma^{}_i = -2\delta^{}_{\alpha\beta}$; 2) $\alpha=\beta=0$, 
$\displaystyle 
{\rm Tr}~\sigma^{}_0\sigma^{}_i\sigma^{}_0\sigma^{}_i = 6$. The vertex is a diagonal matrix 
$\displaystyle \Gamma^{}_{2|\alpha\beta} (0) = \delta^{}_{\alpha\beta} \Gamma_{2|\alpha}(0)$, 
with elements
$$\Gamma^{}_0(0) = -\int\frac{d^3p}{(2\pi)^3}~\frac{\eta^2-p^2}{[p^2+\eta^2]^2},\;\;{\rm  and} \;\;
\Gamma^{}_{i=1,2,3}(0) = -\frac{1}{3}\int\frac{d^3p}{(2\pi)^3}~\frac{\displaystyle 3\eta^2 + p^2}{[\eta^2+p^2]^2}.$$
All elements of the mass matrix are then massive:
\begin{equation}
\label{eq:MassEls3d}
M^{}_0 = \frac{1}{g} - \Gamma^{}_0(0) = 2\int\frac{d^3p}{(2\pi)^3}~\frac{\eta^2}{[p^2+\eta^2]^2} = \frac{\eta}{4\pi}, \;\; 
M^{}_{i=1,2,3} = \frac{1}{g} +  \frac{1}{3}\int\frac{d^3p}{(2\pi)^3}~\frac{\displaystyle 3\eta^2 +p^2}{[\eta^2+p^2]^2} = {\rm\lambda}.
\end{equation}

\section{Details of the perturbative corrections to the DOS}
\label{app:perturbation}

\begin{figure}[t]
\includegraphics[width=4.0cm]{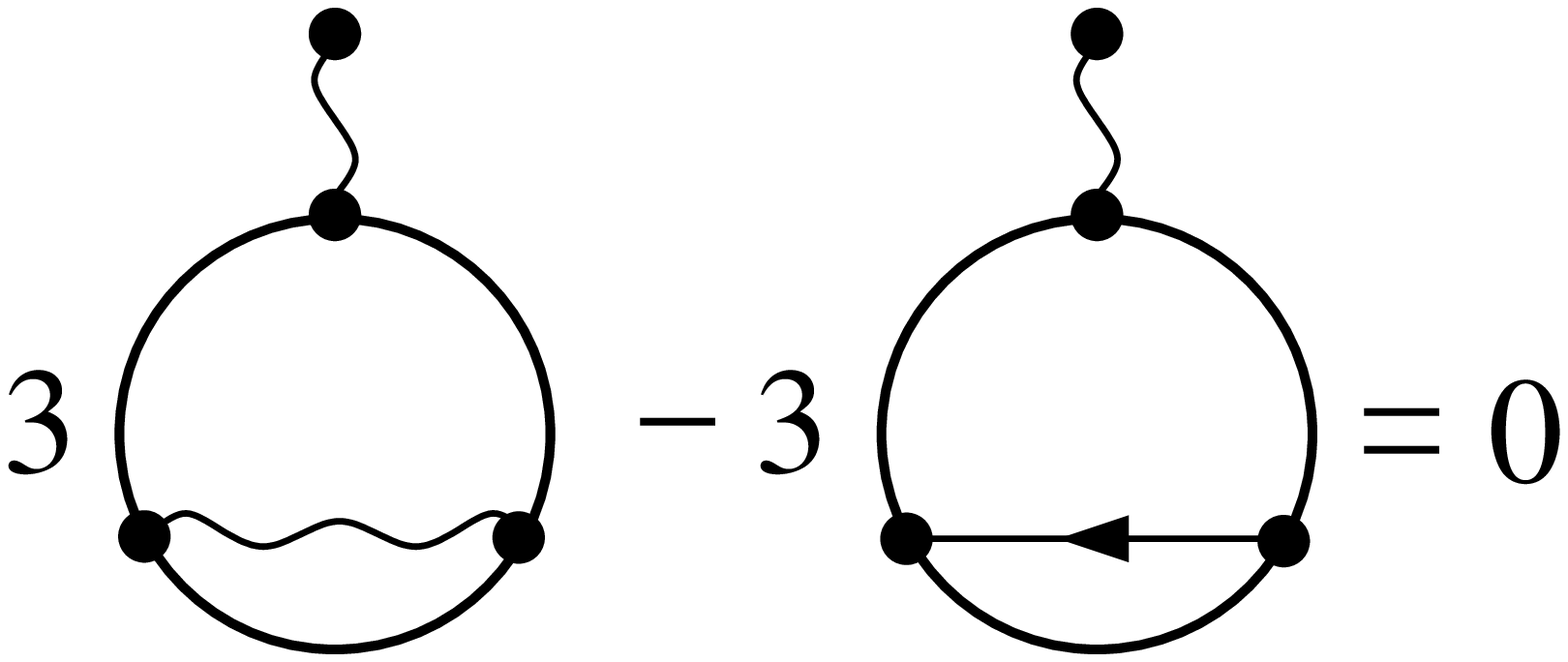}
\caption{Compensation of the leading order corrections from 'rainbow'-diagrams as explained in the main text.
Wavy lines denote contractions of bosonic fields $P$ while straight lines the contractions of Grassmann variables $\bar\chi\chi$. } 
\label{fig:3Ord}
\end{figure}

Main corrections to the DOS are calculated as 
\begin{equation}
\label{eq:Corr1}
\delta G^{ii}_{rr} \sim -i\sqrt{\frac{N}{g}}\sigma^{}_{\nu}\int{\cal D}\hat Q~e^{-{\cal S}^{}_G}~P^{i}_r\left(1 + {\cal S}^{}_p + \cdots  \right)
\ ,
\end{equation}
where ${\cal S}^{}_G$ represents the full Gaussian action and 
\begin{eqnarray} 
{\cal S}^{}_p &=& N \sum_{n\geqslant3}\frac{(-1)^n}{n}\left(\frac{g}{N}\right)^{\frac{n}{2}}{\rm trg}[{\bar G}\hat Q\Sigma^{}_{\nu}]^n,
\end{eqnarray}
where the fields are again rescaled as $\hat Q\to\sqrt{g/N}\hat Q$. It is obvious from Eq.~(\ref{eq:Corr1}) that only terms with an odd power of fields $P$ contribute to the DOS. To order $g^3$ the relevant contributions are 
\begin{eqnarray}
\label{eq:Line1}
{\cal S}^{}_p  \sim -\frac{N}{3}\left(\frac{g}{N}\right)^{\frac{3}{2}}{\rm trg}[{\bar G}\hat Q\Sigma^{}_\nu]^3  
-\frac{N}{5}\left(\frac{g}{N}\right)^{\frac{5}{2}}{\rm trg}[{\bar G}\hat Q\Sigma^{}_\nu]^5. 
\end{eqnarray}
Three-field term becomes after performing the graded trace and retaining only contributions 
with an odd number of $P$'s 
\begin{equation}
\label{eq:Line1a}
-i\frac{N}{3}\left(\frac{g}{N}\right)^{\frac{3}{2}}{\rm tr}\{[{\bar G}P\sigma^{}_\nu]^3 - 3{\bar G}P\sigma^{}_\nu {\bar G}\bar\chi\sigma^{}_\nu {\bar G}\chi\sigma^{}_\nu\},
\end{equation}
and five-field term becomes
\begin{subequations}
\begin{eqnarray}
\label{eq:Line2a}
&&
i\frac{N}{5}\left(\frac{g}{N}\right)^{\frac{5}{2}}{\rm tr}\left\{[{\bar G}P\sigma^{}_\nu]^5 - 5 [{\bar G}P\sigma^{}_\nu]^3{\bar G}\bar\chi\sigma^{}_\nu {\bar G}\chi\sigma^{}_\nu \right\} \\
\label{eq:Line2b}
&+&
iN\left(\frac{g}{N}\right)^{\frac{5}{2}}{\rm tr}\left\{{\bar G}P\sigma^{}_\nu \left({\bar G}\bar\chi\sigma_\nu [{\bar G}Q\sigma^{}_\nu]^2{\bar G}\chi\sigma^{}_\nu 
+[{\bar G}\bar\chi\sigma_\nu {\bar G}\chi\sigma^{}_\nu]^2\right)\right\}. 
\end{eqnarray}
\end{subequations}
Here the fermionic fields are normally ordered in order to guarantee for the positive sign of contractions. 
The contribution from Eq.~(\ref{eq:Line1a}) reads:
\begin{eqnarray}
\label{eq:gsq}
&\displaystyle
%{\cal O}(g^2)
\delta G^{(1)}_{rr} = -\frac{g}{3}\sigma^{}_{\nu}\sigma^{}_a
\sum_{r_1,r_2,r_3}\left[\langle P^{a}_rP^\alpha_{r_1}P^\beta_{r_2}P^\gamma_{r_3}\rangle^{}_{G} 
- 3 \langle P^a_r P^\alpha_{r_1}\rangle^{}_{G} 
\langle\bar\chi^\beta_{r_2}\chi^\gamma_{r_3}\rangle^{}_{G} \right]
\Gamma^{(\nu)}_{3|\alpha r_1,\beta r_2,\gamma r_3},
&
\end{eqnarray}
where the contraction brackets represent functional integration over the Gaussian action. The third order virtual fermion loop reads
\begin{eqnarray}
\label{eq:3FermiLoop}
\Gamma^{(\nu)}_{3|\alpha r_1,\beta r_2,\gamma r_3}={\rm Tr}~
\sigma^{}_{\alpha}\sigma^{}_\nu {\bar G}^{}_{r_1 r_2} 
\sigma^{}_{\beta}\sigma^{}_\nu {\bar G}^{}_{r_2 r_3} \sigma^{}_{\gamma}\sigma^{}_\nu {\bar G}^{}_{r_3 r_1},
\end{eqnarray}
and is invariant under cyclic index permutations. Because of this cyclicity, all three pairwise contractions of fields $P$ contribute equally after index relabeling
\begin{eqnarray}
\langle P^{a}_rP^\alpha_{r_1}P^\beta_{r_2}P^\gamma_{r_3}\rangle^{}_{G} = 
3\langle P^{a}_rP^\alpha_{r_1}\rangle^{}_{G}\langle P^\beta_{r_2}P^\gamma_{r_3}\rangle^{}_{G},
\end{eqnarray}
and since bosonic and fermionic correlators are the same this DOS correction vanishes as a whole. Diagrammatically, this equation is shown in Fig.~\ref{fig:3Ord}. This result is nothing but the manifestation of the linked-cluster theorem and has a very simple meaning, namely it postulates the vanishing of the leading order 'rainbow'-like corrections which are already accounted in the saddle-point equation.

\begin{figure}[t]
\includegraphics[width=7.0cm]{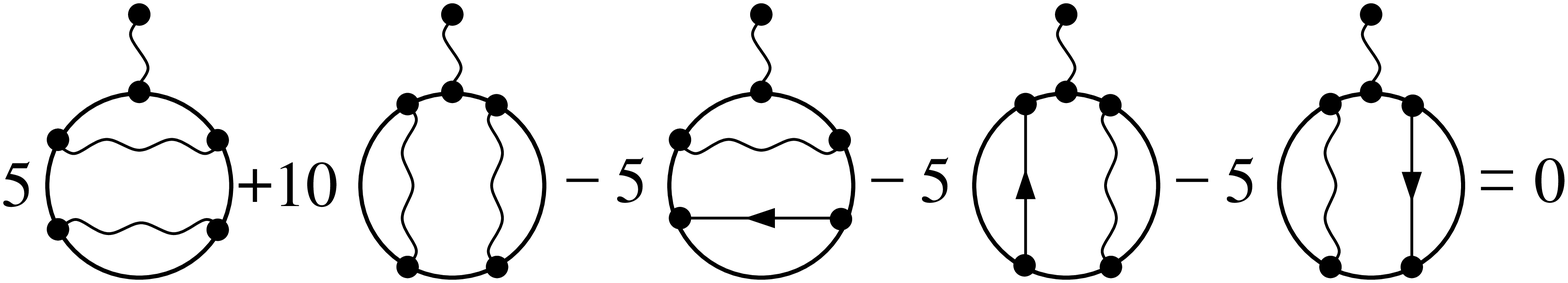}
\caption{Partial compensation of the second order 'rainbow'- and 'bulge'- corrections arising from Eq.~(\ref{eq:Line2a}) as explained in the main text.} 
\label{fig:5Orda}
\end{figure}
\begin{figure}[t]
\includegraphics[width=4.0cm]{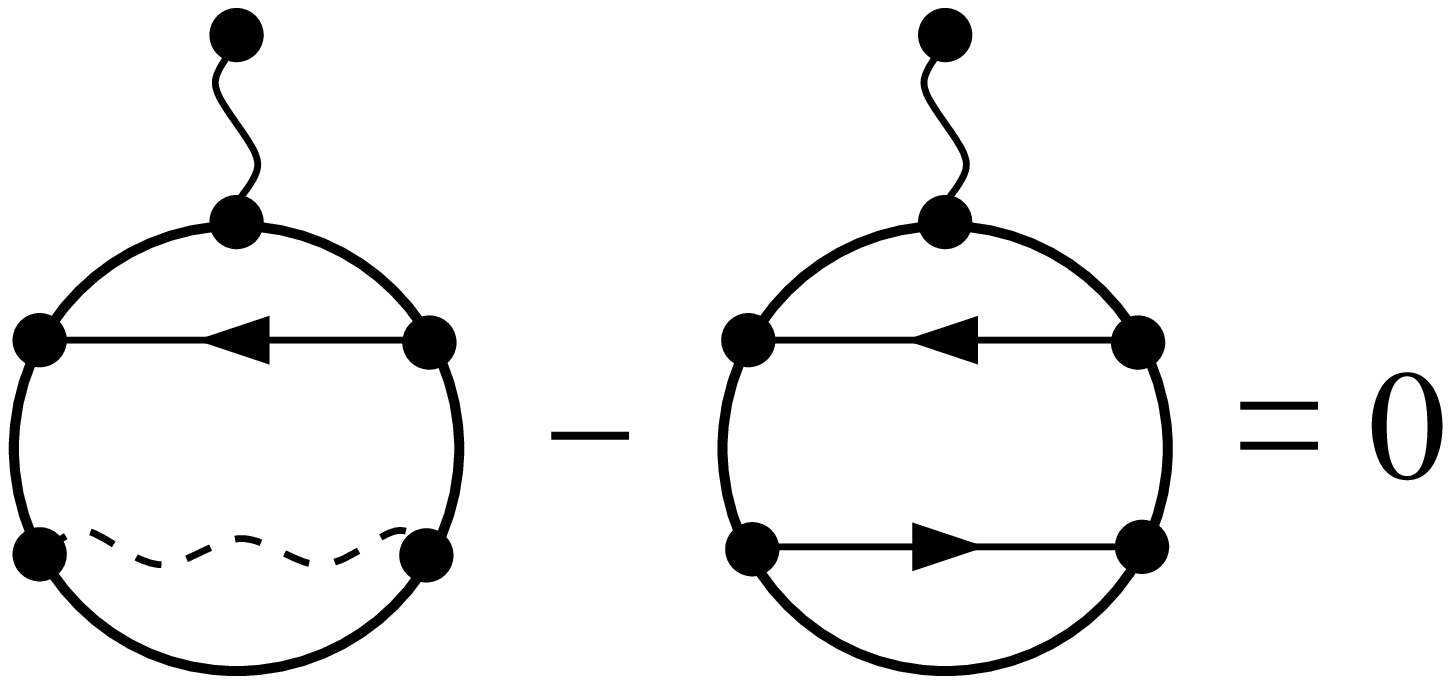}
\caption{Partial compensation of the second order 'rainbow'- corrections arising from Eq.~(\ref{eq:Line2b}) as explained in the main text. The dashed wavy line denotes the contraction of the fields $Q$. Together with Fig.~\ref{fig:5Ordb} the total annihilation of the 'rainbow'-corrections, already included in the saddle-point equation is insured. %The negative sign in front of the second diagram is due to the odd number of Grassmannian permutations necessary for the contractions. 
} 
\label{fig:5Ordb}
\end{figure}

A similar line of reasoning reveals the mutual annihilation of all 'rainbow'- and 'bulge'-like DOS corrections to order $g^2$ encoded in Eq.~(\ref{eq:Line2a}) as depicted in Fig.~\ref{fig:5Orda}. Five non-vanishing pairwise contractions 
\begin{eqnarray}
&\displaystyle
\langle P^a_{r} P^\alpha_{r_1} P^\beta_{r_2} P^\gamma_{r_3} P^\iota_{r_4} P^\tau_{r_5} \rangle^{}_G  
= 5 
\langle P^a_{r} P^\alpha_{r_1} \rangle^{}_G \langle P^\beta_{r_2}P^\iota_{r_4} \rangle^{}_G  \langle  P^\gamma_{r_3} P^\tau_{r_5} \rangle^{}_G 
&
\end{eqnarray}
generate the so-called 'sunrise'-diagrams shown in Fig.~\ref{fig:5Ordc} on the left.  The factorization of the four-fermion term from Eq.~(\ref{eq:Line2b}) is not unique and yields two contributions
\begin{eqnarray}
\label{eq:ChiContr}
\langle P^a_rP^\alpha_{r_1}\bar\chi^\beta_{r_2}\chi^\gamma_{r_3}\bar\chi^\iota_{r_4}\chi^\tau_{r_5}\rangle^{}_{G}  
= \langle P^a_rP^\alpha_{r_1} \rangle^{}_{G} \langle \bar\chi^\beta_{r_2}\chi^\gamma_{r_3}\rangle^{}_{G}\langle  \bar\chi^\iota_{r_4}\chi^\tau_{r_5}\rangle^{}_{G}  
- \langle P^a_rP^\alpha_{r_1} \rangle^{}_{G} \langle \bar\chi^\beta_{r_2}\chi^\tau_{r_5}\rangle^{}_{G}
\langle \bar\chi^\iota_{r_4}\chi^\gamma_{r_3}\rangle^{}_{G},
\end{eqnarray}
where the minus sign in front of the second term is due to the odd number of Grassmannian permutations. Because of the sub-lying supersymmetry, this negative term gets totally annihilated by the term
\begin{eqnarray}
\langle P^a_r P^\alpha_{r_1}\bar\chi^\beta_{r_2} Q^\gamma_{r_3} Q^\iota_{r_4}\chi^\tau_{r_5} \rangle^{}_{G} = 
\langle P^a_rP^\alpha_{r_1} \rangle^{}_{G} \langle \bar\chi^\beta_{r_2}\chi^\tau_{r_5}\rangle^{}_{G}
\langle Q^\iota_{r_4}Q^\gamma_{r_3}\rangle^{}_{G}.
\end{eqnarray}
This is shown diagrammatically in Fig.~\ref{fig:5Ordb}. Hence, the 'rainbow'-like contributions get annihilated to this order too. The positive contraction from Eq.~(\ref{eq:ChiContr}) gives rise to the non-vanishing correction to the DOS in form of a 'bulge'-diagram, shown in Fig.~\ref{fig:5Ordc} on the right. This is a rather as remarkable as unexpected result, since naively diagrams of that type are considered to be one particle reducible. This misapprehension roots in the formal similarity of this diagrammatic approach to that of the non-local self-energy of interacting systems which employes a slightly different version of the linked-cluster theorem. The non-vanishing terms to order $g^2$ are 
\begin{eqnarray}
\nn
\delta G^{(2)}_{rr} &=& \frac{g^2}{N} \sigma^{}_{\nu}\sigma^{}_a \sum_{r_1r_2r_3r_4r_5} 
\Gamma^{(\nu)}_{5|\alpha r_1,\beta r_2,\gamma r_3,\iota r_4,\tau r_5} 
\\
\label{eq:Corr}
&\times &\left[\langle P^a_rP^\alpha_{r_1} \rangle^{}_{G} \langle P^\beta_{r_2}P^\iota_{r_4}\rangle^{}_{G} \langle P^\gamma_{r_3}P^\tau_{r_5}\rangle^{}_{G} 
+ \langle P^a_rP^\alpha_{r_1} \rangle^{}_{G} \langle \bar\chi^\beta_{r_2}\chi^\gamma_{r_3}\rangle^{}_{G}\langle\bar\chi^\iota_{r_4}\chi^\tau_{r_5}\rangle^{}_{G}\right] ,
\end{eqnarray}
with the fifth order virtual fermion loop
\begin{eqnarray}
\label{eq:5FermiLoop}
\Gamma^{(\nu)}_{5|\alpha r_1,\beta r_2,\gamma r_3,\iota r_4,\tau r_5}={\rm Tr}~
\sigma^{}_{\alpha}\sigma^{}_\nu {\bar G}^{}_{r_1 r_2}\sigma^{}_{\beta}\sigma^{}_\nu 
{\bar G}^{}_{r_2 r_3} \sigma^{}_{\gamma}\sigma^{}_\nu {\bar G}^{}_{r_3 r_4} 
\sigma^{}_{\iota}\sigma^{}_\nu {\bar G}^{}_{r_4 r_5} \sigma^{}_{\tau}\sigma^{}_\nu {\bar G}^{}_{r_5 r_1}.
\end{eqnarray}
In ultra-weak limit the correlators are replaced by delta-functions. The detailed evaluation of this correction in ultra-weak disorder limit is given in Appendix~\ref{app:PTloop}. 

\section{Evaluation of the perturbative corrections: Weak disorder limit}
\label{app:PTloop}

In ultra-weak disorder limit the correction to the Green's function which arise from diagrams depicted in Fig.~\ref{fig:5Ordc} read
\begin{eqnarray}
\delta G^{(2)}_{rr} &\sim& \frac{g^2}{N}\sigma^{}_0 {\rm Tr} \sum_{r_1,r_2}\left[
{\bar G}^{}_{r r_1}{\bar G}^{}_{r_1 r_1}{\bar G}^{}_{r_1 r_2}{\bar G}^{}_{r_2 r_2}{\bar G}^{}_{r_2 r} +
{\bar G}^{}_{r r_1}{\bar G}^{}_{r_1 r_2}{\bar G}^{}_{r_2 r_1}{\bar G}^{}_{r_1 r_2}{\bar G}^{}_{r_2 r}
\right].
\end{eqnarray}
First term is harmless: each of the two 'bulges' can be expressed using the saddle-point condition as
${\bar G}^{}_{rr} = -i{\eta}/{g}$, while the remaining loop converges in both dimensions. 
\begin{equation}
\label{eq:bulges}
{\cal D}^{}_1 = 
{\rm Tr} \sum_{r_1,r_2} {\bar G}^{}_{r r_1}{\bar G}^{}_{r_1 r_1}{\bar G}^{}_{r_1 r_2}{\bar G}^{}_{r_2 r_2}{\bar G}^{}_{r_2 r} = 
-\frac{\eta^2}{g^2}{\rm Tr}\int\frac{d^dq}{(2\pi)^d}~{\bar G}^3(q) = 2i\frac{\eta^3}{g^2}\int\frac{d^dq}{(2\pi)^d}~\frac{3q^2-\eta^3}{[q^2+\eta^2]^3}, 
\end{equation}
which leads to
\begin{equation}
{\cal D}^{}_1 = \frac{i}{2\pi}\frac{\eta^{d-1}}{g^2}.
\end{equation}

To the contrary, the evaluation of the first contribution is technically more demanding. Transforming the loop into the Fourier space
we get
\begin{equation}
{\cal D}^{}_2 = {\rm Tr} \sum_{r_1,r_2} {\bar G}^{}_{r r_1}{\bar G}^{}_{r_1 r_2}{\bar G}^{}_{r_2 r_1}{\bar G}^{}_{r_1 r_2}{\bar G}^{}_{r_2 r} = 
 {\rm Tr}\int\frac{d^dp d^dq}{(2\pi)^{2d}}~{\bar G}(q+p){\bar G}(q)\int\frac{d^dk}{(2\pi)^d}~{\bar G}(k-p){\bar G}(k){\bar G}(k). 
\end{equation}
Integrals over $q$ and $k$ can be carried out separately using Feynman representation of the fraction product:
\begin{equation}
\label{eq:Feynman} 
\frac{1}{A^{1+n}B} = \int_0^1 dx\frac{(n+1)(1-x)^n}{[(1-x)A+xB]^{2+n}}.
\end{equation}
The $q$-integral reads
\begin{equation}
I^{}_1 = 
\int\frac{d^dq}{(2\pi)^d}~{\bar G}(q+p){\bar G}(q) = \int^1_0 dx\int\frac{d^dq}{(2\pi)^d}~\frac{[\slashed q+\slashed p-i\eta][\slashed q -i\eta]}
{[x(q+p)^2+(1-x)q^2+\eta^2]^2}.
\end{equation}
By shifting $q_i\to q_i-xp_i$ the denominator becomes rotationally invariant, which enables us to drop all odd powers of $q_i$ in the numerator, getting
\begin{equation}
I^{}_1 = \int^1_0 dx\int\frac{d^dq}{(2\pi)^d}~\frac{q^2-\eta^2-x(1-x)p^2-i\eta\slashed p(1-2x)}{[q^2+\eta^2+x(1-x)p^2]^2}.
\end{equation}
One recognizes that the term with $1-2x$ vanishes after integration over $x$: Since $\displaystyle 1-2x=\frac{d}{dx}x(1-x)$ and the remaining expression depends only on $x(1-x)$ we get
\begin{equation}
\int_0^1 dx~ f[x(1-x)] \frac{d}{dx}[x(1-x)] = \int_0^1 dx~\frac{d}{dx}F[x(1-x)] = F[0]-F[0] = 0,
\end{equation}
where $F(x)$ is the indefinite integral of $f(x)$. Since hence $I^{}_1$ is symmetric under $p\to-p$ we can omit all odd powers of $p$ in the  integral over $k$. In 2d the remaining integral $I^{}_1$ was computed in~[\onlinecite{PhysRevB89}]. We get assuming a very large upper cutoff and using the saddle-point condition  
\begin{equation}
\label{eq:FirInt}
I^{\rm 2d}_1  \sim \frac{1}{g} - 
\frac{1}{2\pi}\left.\sqrt{\frac{4+t^2}{t^2}}{\rm atanh}\sqrt{\frac{t^2}{4+t^2}}\right|_{t=p/\eta}. 
\end{equation}
The evaluation in 3d takes a few computational lines more: Splitting the integrand in divergent and convergent parts
\begin{equation}
I^{\rm 3d}_1 = \int_0^1 dx\int\frac{d^3q}{(2\pi)^3}\left[\frac{1}{q^2+\eta^2+x(1-x)p^2} -2 \frac{\eta^2+x(1-x)p^2}{[q^2+\eta^2+x(1-x)p^2]^2} \right]
\end{equation}
we can perform $q$-integral in the convergent part. We continue by adding and subtracting $1/g$ to the divergent part and using the saddle point equation:
\begin{equation}
I^{\rm 3d}_1 = \frac{1}{g}-\frac{1}{4\pi}\int_0^1dx\sqrt{\eta^2+x(1-x)p^2} + \frac{1}{2\pi^2}\int_0^1dx\int_0^\infty dq~\left[\frac{q^2}{q^2+\eta^2+x(1-x)p^2} -
\frac{q^2}{q^2+\eta^2} \right].
\end{equation}
The divergent contribution in the remaining $q$-integral cancels, hence the integral can be carried out using the residue theorem: 
\begin{equation}
I^{\rm 3d}_1 = \frac{1}{g}+\frac{\eta}{4\pi} - \frac{\eta}{2\pi}\left.\int_0^1 dx \sqrt{1+x(1-x)t}\right|_{t=p/\eta}.
\end{equation}
The indefinite integral over $x$ is known, putting the boundaries and simplifying the expression we finally get
\begin{equation}
I^{\rm 3d}_{1} \sim \frac{1}{g} - \frac{\eta}{8\pi}\left. \frac{4+t^2}{t}{\rm atan}\left(\frac{t}{2}\right)\right|_{t=p/\eta}.
\end{equation}

Second integral can be evaluated in a similar fashion: Using the Feynman parametrization we get
\begin{eqnarray}
I^{}_2 = \int\frac{d^dq}{(2\pi)^d} ~ {\bar G}(q-p){\bar G}(q){\bar G}(q) &=& 2\int_0^1dx(1-x)\int \frac{d^dq}{(2\pi)^d}~\frac{[\slashed q-\slashed p-i\eta][\slashed q-i\eta][\slashed q-i\eta]}{[(1-x)q^2+x(q-p)^2+\eta^2]^3}\\
&=& -2i\eta\int_0^1 dx(1-x)\int\frac{d^dq}{(2\pi)^d}~\frac{3q^2-\eta^2-x(1-x)p^2}{[q^2+\eta^2+x(1-x)p^2]^3}.
\end{eqnarray}
Power counting indicates that the integral over $q$ converges in both dimensions. The symmetrization of the denominator is achieved by shifting $q_i\to q_i+xp_i$, when we dropped odd powers of $q$ and $p$ and regrouped $x$-dependent factors at $p^2$ using the fact that the integral operator $\int_0^1 dx$ does not change under substitution $x\to1-x$. This leads in 2d to
\begin{eqnarray}
\label{eq:SecInt2d}
I^{\rm 2d}_2 = -\frac{i}{2\pi\eta}\left.\int_0^1dx~\frac{1-x}{1+x(1-x)t^2}\right|_{t=p/\eta} = 
-\frac{i}{\pi\eta}\left.\frac{1}{\sqrt{t^2(4+t^2)}}~{\rm atanh}\sqrt{\frac{t^2}{4+t^2}}\right|_{t=p/\eta}.
\end{eqnarray}
In 3d we analogously get
\begin{eqnarray}
\label{eq:SecInt3d}
I^{\rm 3d}_2 = -\frac{i}{2\pi}\left. \int_0^1dx\frac{1-x}{\sqrt{1+x(1-x)t^2}}  \right|_{t=p/\eta} = -\frac{i}{2\pi t}\left.{\rm atan}
\left(\frac{t}{2}\right) \right|_{t=p/\eta}.
\end{eqnarray}
Taking the trace over the Dirac space becomes trivial and gives a factor 2. In 2d we obtain with Eqs.~(\ref{eq:FirInt}) and (\ref{eq:SecInt2d}) 
\begin{equation}
{\cal D}^{\rm 2d}_2 =  
-2i\frac{\eta}{\pi}\int\frac{d^2t}{(2\pi)^2}~\frac{{\rm atanh}\sqrt{\frac{t^2}{t^2+4}} }{t\sqrt{t^2+4}} 
\left[\frac{1}{g} - \frac{1}{2\pi}\sqrt{\frac{t^2+4}{t^2}}{\rm atanh}\sqrt{\frac{t^2}{t^2+4}} \right].
\end{equation}
Extracting from the saddle-point equation Eq.~(\ref{eq:SPC}) the fitting expression 
\begin{equation}
\frac{1}{2}\log\left[1 + \frac{\Lambda^2}{\eta^2}\right] = f^{}_{\rm 2d} \equiv \frac{2\pi}{g},
\end{equation}
we obtain with high accuracy 
\begin{eqnarray}
-i\frac{\eta}{\pi^2 g}\int_0^{\Lambda/\eta}dt~\frac{{\rm atanh}\sqrt{\frac{t^2}{t^2+4}} }{\sqrt{t^2+4}} &\sim& 
-i\frac{1}{\pi^2}\frac{\eta}{g}\frac{f^2_{2d} }{2}=-i2\frac{\eta}{g^3},\\
\nn
i\frac{4\eta}{(2\pi)^3} \int_0^{\Lambda/\eta}dt~\frac{1}{t}\left[{\rm atanh}\sqrt{\frac{t^2}{t^2+4}} \right]^2 &\sim& 
i\frac{4\eta}{(2\pi)^3}\left[\frac{1}{2} + \frac{f^3_{\rm 2d}}{3}\right] 
= i\frac{\eta}{g}\left[
\frac{4}{3g^2} + \frac{2g}{(2\pi)^3}
\right],
\end{eqnarray}
where in the second equality the saddle-point condition is used. Counting ${\cal D}^{\rm 2d}_1$ and ${\cal D}^{\rm 2d}_2$ together we finally get the correction to the  
\begin{equation}
\delta G^{ii}_{rr} \sim -i\frac{\varrho^{}_{SCBA}}{2N^2}\left[\frac{2}{3} - \frac{g}{2\pi} - \frac{g^3}{4\pi^3} \right]\sigma^{}_0,
\end{equation}
and from here the DOS correction given in Eq.~(\ref{eq:corr_2d}). In 3d, the remaining integral reads 
\begin{eqnarray}
\nn
{\cal D}^{\rm 3d}_2 &=& -\frac{i\eta^3}{g\pi} \int\frac{d^3t}{(2\pi)^3} \frac{1}{t}{\rm atan}\left(\frac{t}{2}\right) + 
\frac{i\eta^4}{8\pi^2}\int\frac{d^3t}{(2\pi)^3}\frac{4+t^2}{t^2}\left[{\rm atan}\left(\frac{t}{2}\right)\right]^2 \\
\label{eq:3dScL}
&=& -\frac{i}{g}\frac{\eta^3}{2\pi^3}\int_0^{\Lambda/\eta}dt~ t~{\rm atan}\left(\frac{t}{2}\right) + 
\frac{i\eta^4}{(2\pi)^4}\int_0^{\Lambda/\eta}dt~(4+t^2)\left[{\rm atan}\left(\frac{t}{2}\right)\right]^2.
\end{eqnarray}
From the saddle-point equation Eq.~(\ref{eq:SPC}) we get the fitting polynomial
\begin{equation}
\frac{\Lambda}{\eta} - {\rm atan}\left(\frac{\Lambda}{\eta}\right) = f^{}_{\rm 3d} \equiv \frac{2\pi^2}{g\eta}. 
\end{equation}
Fitting integrals in Eq.~(\ref{eq:3dScL}) with different powers of the polynomial $f^{}_{\rm 3d}$ we obtain with excellent accuracy
\begin{equation}
I^{\rm 3d}_2 \sim -\frac{i}{g}\frac{\eta^3}{2\pi^3}\frac{\pi}{4}f^2_{\rm 3d} + \frac{i\eta^4}{(2\pi)^4}\left( \pi^2 f^{}_{\rm 3d} + \frac{5\pi}{19}f^3_{\rm 3d} \right) 
= -i\frac{\eta}{g}\left[\frac{\pi^2}{2g^2}\left(1-\frac{5\pi}{19} \right) - \frac{\eta^2}{8} \right].
\end{equation}
Counting  ${\cal D}^{\rm 3d}_1$ and ${\cal D}^{\rm 3d}_2$ together we eventually obtain
\begin{equation}
\delta G^{ii}_{rr} \sim -i\frac{\varrho^{}_{SCBA}}{2N^2} \left[\frac{\pi^2}{2}\left(1 - \frac{5\pi}{19}\right) - \frac{g\eta}{2\pi} 
- \frac{g^2\eta^2}{8}\right]\sigma^{}_0,
\end{equation}
which upon taking the trace over the Dirac space and  the imaginary part yields the correction in Eq.~(\ref{eq:corr_3d}).

\section{Evaluation of the perturbative corrections: Strong disorder limit}
\label{app:PTloopSt}
Here we get 
\begin{eqnarray}
\delta G^{(2)}_{rr} &\sim& \frac{\pi g}{16N}\sigma^{}_0 {\rm Tr} \sum_{r_1,r_2}\left[
{\bar G}^{}_{r r_1} \sigma^{}_3 {\bar G}^{}_{r_1 r_1} \sigma^{}_3 {\bar G}^{}_{r_1 r_2} \sigma^{}_3 {\bar G}^{}_{r_2 r_2} \sigma^{}_3 {\bar G}^{}_{r_2 r} +
{\bar G}^{}_{r r_1} \sigma^{}_3 {\bar G}^{}_{r_1 r_2} \sigma^{}_3 {\bar G}^{}_{r_2 r_1} \sigma^{}_3 {\bar G}^{}_{r_1 r_2} \sigma^{}_3 {\bar G}^{}_{r_2 r}
\right].
\end{eqnarray}
The evaluation of the first contribution is entirely analogous to the weak disorder case, we get
\begin{eqnarray}
{\cal D}^{}_1 = 
\frac{\pi g}{16N} {\rm Tr} \sum_{r_1,r_2}
{\bar G}^{}_{r r_1} \sigma^{}_3 {\bar G}^{}_{r_1 r_1} \sigma^{}_3 {\bar G}^{}_{r_1 r_2} \sigma^{}_3 {\bar G}^{}_{r_2 r_2} \sigma^{}_3 {\bar G}^{}_{r_2 r} 
= i\frac{\varrho^{}_{SCBA}}{(8N)^2}.
\end{eqnarray}
Second contribution reads
\begin{equation}
{\cal D}^{}_2 = 
 {\rm Tr}\int\frac{d^2p d^2q}{(2\pi)^{4}}~{\bar G}(q)\sigma^{}_3{\bar G}(q+p)\sigma^{}_3\int\frac{d^2k}{(2\pi)^2}~{\bar G}(k+p)\sigma^{}_3{\bar G}(k){\bar G}(k)\sigma^{}_3. 
\end{equation}
The presence of the $\sigma^{}_3$ matrix which anticommutes with the Dirac Hamiltonian changes the sign of the $q$-integral:
\begin{eqnarray}
\nn
\int\frac{d^2q}{(2\pi)^{2}}~{\bar G}(q)\sigma^{}_3{\bar G}(q+p)\sigma^{}_3 &=& 
-\left(\frac{1}{g} -\frac{1}{2\pi}\frac{t}{\sqrt{4+t^2}}{\rm atanh}\sqrt{\frac{t^2}{4+t^2}} \right) \\
&&+i\frac{\slashed t}{\pi}\frac{1}{t\sqrt{4+t^2}}{\rm atanh}\sqrt{\frac{t^2}{4+t^2}},
\end{eqnarray}
where again $t=p/\eta$ and $\slashed t=t^{}_i\sigma^{}_{i=1,2}$. Second integral becomes 
\begin{eqnarray}
\int\frac{d^2k}{(2\pi)^2}~{\bar G}(k+p)\sigma^{}_3{\bar G}(k){\bar G}(k)\sigma^{}_3 = \frac{2i-\slashed t}{2\pi\eta}
\left(\frac{1}{4+t^2} - \frac{t}{(4+t^2)^{3/2}}{\rm atanh}\sqrt{\frac{t^2}{4+t^2}} \right), 
\end{eqnarray}
which eventually leads to 
\begin{equation}
{\cal D}^{}_2 = -i\frac{\varrho^{}_{SCBA}}{2\pi N} \int_0^{\Lambda/\eta}dt~\left(\frac{t}{4+t^2} - \frac{t^2}{(4+t^2)^{3/2}}{\rm atanh}\sqrt{\frac{t^2}{4+t^2}} \right) =
-i\frac{\varrho^{}_{SCBA}}{2\pi N} \left[\frac{2\pi}{g} - \frac{1}{2}\left(\frac{2\pi}{g}\right)^2 \right].
\end{equation}
The integrals can be evaluated analytically. Adding the contributions from all diagrams and extracting the DOS we finally get
\begin{equation}
 \varrho \sim \varrho^{}_{SCBA} + \frac{\varrho^{}_{SCBA}}{(4N)^2}\left(\frac{3}{2} - \frac{2\pi}{g}\right).
\end{equation}

\end{widetext}

\end{document}